\documentstyle[sprocl,epsfig]{article}

\def\NP{{\em Nucl.\ Phys.\ }}
\def\PL{{\em Phys.\ Lett.\ }}
\def\PRL{{\em Phys.\ Rev.\ Lett.\ }}
\def\PR{{\em Phys.\ Rev.}}

\def\etal{{\it et al}.}

\def\lamF{\lambda_F}
\def\lamG{\lambda_G}
\def\mrec{m_{rec}}

\def\pra{|{\rm PRA}|}
\def\fx{F_x}
\def\gx{G_x}
\def\uglu{\hskip 0pt plus 1fil minus 1fil}
\def\uglux{\hskip 0pt plus .75fil minus .75fil}
\def\slashed#1{\setbox200=\hbox{$ #1 $}
    \hbox{\box200 \hskip -\wd200 \hbox to \wd200 {\uglu $/$ \uglux}}}

\def\be{\begin{equation}}
\def\ee{\end{equation}}
\def\bea{\begin{eqnarray}}
\def\eea{\end{eqnarray}}

\newcommand{\lsim}{\mathrel{\lower4pt\hbox{$\sim$}}
\hskip-12.5pt\raise1.6pt\hbox{$<$}\;}

\newcommand{\gsim}{\mathrel{\lower4pt\hbox{$\sim$}}
\hskip-12.5pt\raise1.6pt\hbox{$>$}\;}

\begin{document}
%
%
%
%%%%%%%%%%%%%%%%%%%%%%%%%%%%%%%%%%%%%%%
% comment the below out to remove stamp
%%%%%%%%%%%%%%%%%%%%%%%%%%%%%%%%%%%%%%%
%
%

\title{CLOSING IN ON $B$-CP}

\author{David Atwood}

\address{Department of Physics and Astronomy, Iowa 
State University, \\ Ames, IA\ \ 50011\\
Email: atwood@iastate.edu}

\author{Amarjit Soni}

\address{Theory Group, Brookhaven National Laboratory, \\ Upton, NY\ \ 11973\\
E-mail: soni@penguin.phy.bnl.gov}


\maketitle\abstracts{ Major new experimental efforts on 
detecting CP violation in $B$ decays will very soon go on the air. Recent
developments suggest that final state interaction phases in exclusive
decays of $B$ are unlikely to be small indicating the possibility of
observable 
direct CP asymmetries in these channels. CLEO results on charmless
hadronic modes suggest penguin amplitudes are rather big implying that the
extraction of $\alpha$ from $B^0$-$\bar B^0$ alone will be difficult,
thus necessitating also information from direct CP\null. Importance of
$B(B_s)$ decays to two vectors for model independent tests of
electroweak penguins and for extraction of $\alpha(\gamma)$ is
emphasized. Inclusive $b\to sg^\ast$ and related modes, e.g.\ $B\to
\eta^\prime X_s$, are very good probes of CP-odd phase(s) due to beyond
the standard model physics. On the other hand, $b\to dg^\ast$ and
related modes, e.g.\ $B\to \eta^\prime X_d$ are more suitable for CP
violation due to the CKM phase. Two body $b$-quark decays: $b\to Mq_f$
leading to semi-inclusive, $B$ decays, $B\to MX$ (with $2\lsim
E_M\lsim2.8$ GeV), are very interesting and important. Their theory is
relatively clean; partial rate asymmetries may be large in several
cases (e.g.\ $M=K^\ast, K, \rho, \pi\dots$) and a few cases provide
very good probe of electroweak penguins.}

\section{Introduction and Summary\protect\cite{talks}}

Following is the outline of this talk:

\renewcommand{\theenumi}{\Roman{enumi}}

\begin{enumerate}

\item Introduction and Summary

\item CLEO97: A robust penguin $\Rightarrow B^0$-$\bar B^0$ cannot do it
alone. 

\item In exclusive $B$-decays, final state rescattering phases are
unlikely to be small. 

\item Repercussions of long distance rescattering for direct CP.

\item $B\to V_1V_2$: Model independent tests for electroweak penguins
and extraction of the angles of the unitarity triangle. 

\item Chasing LUCY with beauty: Search for a non-standard-phase  via
$b\to sg^\ast$, $B\to \eta^\prime X_s$ and related modes. 

\item $b\to dg^\ast$, $B\to \eta^\prime X_d$ and related modes for
direct CP violation due SM\null. 

\item Two body $b$ decays: $b\to qM$ for direct CP studies and for
looking for electroweak penguins. 

\end{enumerate}

\renewcommand{\theenumi}{\arabic{enumi}}

The asymmetric $B$-factories at KEK and SLAC and the symmetric one at
CORNELL will begin operation in about six months. There is great degree
of expectation that these will soon provide us important new clues with
regard to understanding an intriguing mystery of nature---CP violation. 

Although the centerpiece of the asymmetric $B$-factories is the study of
mixing induced CP via measurements of time dependent asymmetries in
$B^0$-$\bar B^0$ decays, it is becoming increasingly clear that a precise
determination of the angles of the unitarity triangle, essential for
testing the SM, will require a broader understanding of CP violation
i.e.\ involving extensive studies of direct CP violation dealing with
charged as well as neutral $B$ mesons. 

The main point is that recent observation of the charmless, hadronic,
exclusive modes by CLEO strongly indicate that the size of the penguin
amplitudes is appreciable.\cite{linguel,godang,bergfeld} This renders
the extraction of $\alpha$ or $\gamma$ through $B^0$-$\bar B^0$
observations alone rather difficult. Subtraction of the penguin
amplitudes will require study of CP asymmetries in charged $B$ decays as
well. 

An essential ingredient that drives direct CP violation is the final
state (FS) rescattering (CP-even) phase. Reliable calculation of this phase
from theory still remains and outstanding and important challenge. Some
phenomenological arguments as well as indirect indications suggest that
these phases may not be small at the $B$ mass. Thus sizeable direct CP
violations in $B$-decays are very likely. Indeed, the FS
rescattering phases in exclusive channels originate at least in part due
to soft-physics (i.e.\ non-perturbative effects) and can lead to a
different pattern of CP asymmetries\cite{atwood1} than hitherto
envisioned to emerge 
from short distance operators. In exclusive modes (such as
$K\pi$) appreciable asymmetries are plausible.\cite{atwood1}

Direct CP measurements in some of the exclusive channels are not
entirely in vain insofar as determination of the unitarity angle goes.
For example, measurements in the $K\pi$ can be used for quantitatively
testing for the presence of electroweak penguins (EWP).\cite{atwood1} If EWP contributions
to $K\pi$ are found to be negligible then $B^0$, $\bar B^0$, $B^\pm\to
K\pi$ studies can be used to determine $\alpha$.\cite{nir}

$B$, $B_s$ decays to two vector mesons can be especially useful in this
context.\cite{atwood2} Angular distribution of the decays of the vector
mesons provide information on
polarization of the vector mesons.\cite{valen} Model independent tests
can be constructed to monitor the presence of EWP\null.\cite{atwood2}
Modes that show negligible contribution from EWP can then be used to
extract $\alpha$ and $\gamma$.\cite{atwood2}

Inclusive $b\to s$ transitions are a very good probe for non-standard CP
violation phases.\cite{atwood3,hou1} Of particular interest is $b\to
sg^\ast$ since it has $Br\sim1\%$.\cite{hou2} However, as such this mode
is very difficult to detect. CLEO has recently reported a large signal
for $B\to \eta^\prime X_s$.\cite{browder} Experimental data, to date,
supports the interpretation that an appreciable fraction of the signal
comes from $g^\ast\to g\eta^\prime$ through the anomalous coupling of
the $\eta^\prime$ to the two gluons.\cite{atwood4} Consequently $B\to
\eta^\prime X_s$ becomes a good search for the
non-standard-phase (NSP).\cite{atwood3,hou1} It may be useful to extend
the search to 
other expected fragments $(X_g)$ of $g^\ast\to X_g g$. Good examples of
$X_g$ (in addition to $\eta^\prime$) are $\eta(1440)$, $f_0(980)$,
$f_2(1270)$, $K^+K^-$, $\pi^+\pi^-$\dots\cite{atwood3}

The penguin transition $b\to d$ on interference with the tree
contribution $b\to u\bar u d$ is likely to be a good way to detect
direct CP violation from the SM\null. The modes of interest here are
completely analogous to those in the previous paragraph except for the
replacement of $X_s\to X_d$. 

The last topic to be emphasized in this talk is two body decays of the
$b$ quark: $b\to Mq_f$ where $M$ is spin 0 or 1 meson.\cite{atwood5}
These decays lead to semi-inclusive $B$ decays. Theory for these should
be cleaner than exclusive decays [such as $B\to K\pi$, $\rho\pi$\dots]
and their experimental signatures are rather distinctive. Many of the modes
could have large partial rate asymmetries (PRA). Two of the modes $B\to
\rho^0 X_s$ and $\pi^0 
X_s$ are also a good way to look for EWP as the latter dominate over other
contributions.\cite{atwood5}

\section{CLEO97: A Robust Penguin $\Rightarrow B^0$-$\bar B^0$ Cannot do it Alone}

After years of anticipation, in 1997, CLEO reported the first
observation of some charmless, hadronic decay modes; both exclusive and
inclusive.\cite{linguel,godang,bergfeld,browder,roy} For our purpose, the
most interesting examples of these two categories
are:\cite{godang,browder}

\bea
Br (B\to K^+\pi^-) & = & (1.5\pm .5\pm.1)\times 10^{-5} \label{eq1} \\
Br (B\to \eta^\prime X_s) & = & (6.2\pm1.6 ({\rm stat})\pm 1.3 ({\rm
syst})^{+0.0}_{-1.5} ({\rm bkg})) \times 10^{-4} \label{eq2}
\eea

\noindent In addition, CLEO has reported lack of a statistically
significant signal in the $\pi\pi$ mode:\cite{linguel,roy}

\be 
Br (B\to \pi^+\pi^-) \le (1.5\times10^{-5})\; @90\%{\rm CL} \label{eq3}
\ee

A simple way to understand (\ref{eq1}) and (\ref{eq3}) is to assume that
the $\pi\pi$ mode is dominated by tree. Due to the Cabibbo angle, the
tree contribution to $K\pi$ must therefore be extremely small compared
to (\ref{eq1}). Thus $B\to
K\pi$ with the stated $Br$ is dominated by the penguin. Also the penguin
contribution to $\pi\pi\sim |V_{td}|^2\ast 1.5\times10^{-5}/|V_{ts}|^2\sim
10^{-6}$, assuming $|V_{td}/V_{ts}| \sim \lambda \sim .22,$ which is quite
consistent with the existing constraints. Thus the amplitude ratio of
tree vs penguin for $\pi\pi$ is about:

\be
\frac{P_{\pi\pi}}{T_{\pi\pi}} \gsim .25 \label{eq4}
\ee

Therefore, for a precision determination of the unitarity angle
$\alpha$, it does not appear safe to ignore the penguin contribution to
$\pi\pi$. Assuming EWP contribution is negligible,
extraction of $\alpha$ from $B\to \pi\pi$ is still possible but it
requires observation of direct CP i.e.\ partial rate asymmetry (PRA)
measurements in the $B^\pm$ will also be needed. 
In addition, the $Br$
for $\pi^0\pi^0$, likely to be quite small, will also be required.\cite{gronau} 

In fact CLEO measurements (\ref{eq1}) and (\ref{eq3}) and the resultant
$P_{\pi\pi}/T_{\pi\pi}$ also  means that the penguin contribution for
$B_s\to\rho K^0_s$ is likely to be very large. Indeed, since the tree is
then color suppressed it seems the penguin over tree ratio for $B_s\to
\rho K_s$:

\be 
\frac{P(B_s\to\rho K_s)}{T(B_s\to \rho K_s)} \sim 0(1) \label{eq5}
\ee

\noindent Thus this ``text-book example''\cite{sanda} for extracting
$\gamma$ will also become very difficult to implement even when the
study of 
$B_s$-$\bar B_s$ oscillations become an experimental reality. 

Thus the first CLEO results on charmless $B$-decays are indicating that
studies of $B^0$-$\bar B^0$ for the extraction of angles ($\alpha$ and
$\gamma$) of the 
unitarity triangle are unlikely to suffice; information from direct CP
in $B^\pm$ decays is also likely to play an important role.
Interestingly there are good reasons to suspect that direct CP
asymmetries in many exclusive modes may be appreciably large. 

\section{In Exclusive $B$-Decays Final State Rescattering Phases are
Unlikely to be Small.}

From the outset let us emphasize that a reliable methodology for
calculating final state phases is not available. So, any conclusions
about final state rescattering phases we arrive at will suffer from that
drawback. Despite that reservation, it is still perhaps useful to point
out that there are several reasons to suspect that final states
rescattering phases in exclusive decays of $B$ are unlikely to be
negligibly small:

\begin{enumerate}

\item A long standing problem in $B$-decays is that the measured value
of the semi-leptonic branching ratio is a bit (about 10\%) less than the
number predicted by theory. Most likely this difference between the two
numbers originates from the error in the theoretical calculation of the
hadronic $B$-width. It is quite possible that value of the total
hadronic width calculated by using short distance perturbation theory is
missing a contribution due to soft physics or non-perturbative effects. 

\item The $B$-baryon lifetime is significantly different from that of
$B$-mesons. Short-distance perturbation theory and/or heavy quark
effective field theory apparatus is unable to account for such a
difference. Most likely this originates from spectator quark
interactions which are strongly intertwined with FS interactions. 

\item There is a surprising reversal of trend as one goes from $D$ to
$B$ decays. For $D$'s one has\cite{barnett}

\be 
\Gamma(D^0\to K\pi)/\Gamma (D^+\to K\pi) >1. \label{eq6}
\ee

In contrast, for the $B$'s one finds

\be
\Gamma(B^0\to D\pi)/\Gamma(B^+\to D\pi) <1. \label{eq7}
\ee

\item Donoghue \etal\cite{donoghue} have recently examined the
dependence of the FS rescattering phases on the mass of the decaying
$b$-quark; specifically they deal with the case of the $B\to K\pi$ mode.
Assuming that the total cross-section for $K\pi$ obeys the same scaling
laws as that for the other total hadronic cross sections\cite{donna}
and using the 
optical theorem they show that the effects of elastic rescattering do
not diminish as $m_B$ gets larger.\cite{donoghue} In fact Donoghue
\etal\cite{donoghue} 
claim that the effect of inelastic channels\cite{wolf} is even much bigger than the
elastic one leading them to the conclusion that FS rescattering phases
are unlikely to disappear at $m_B$. 

\end{enumerate}

So, while a compelling theoretical argument is not available, it is
quite plausible that final state rescattering phases in exclusive modes
are appreciable; final word on the subject will have to come from
experiment. 

\section{Repercussions of Long Distance Rescattering for Direct CP in
Exclusive Modes}

\subsection{Possibility of Large Direct CP in Exclusive Modes such as
$K\pi$~~\protect\cite{atwood1} }

FS phases due to long distance (LD) rescattering endow direct CP in
exclusive modes a 
rich structure as can be understood by the use of the CPT theorem.
Recall that the theorem requires that the total lifetime of $B$ and
$\bar B$ to be exactly equal whereas CP symmetry requires the partial
widths of $B$ and $\bar B$ into conjugate modes to be equal. As a
result, when the CP violating partial width difference (PWD) for all the
different modes are added up there has to be an exact cancellation to
render the equality of the lifetimes. Furthermore, assuming isospin is
exactly conserved by the strong interactions, the final states with
different isospins cannot participate in the cancellation of PWD\null. 

If multiparticle inelastic channels composed of light quarks make
important contributions to the rescattering phases of FS such as
$\pi\pi$, $K^\ast\pi$ etc.\ then the PWD in these FS must cancel against
the PWD occurring amongst those multiparticle states responsible for
the phase difference,  as required by the
CPT theorem. In other words, the cancellation of the PWD for FS such as
$\pi\pi$, $K^\ast\pi$ etc.\ need not and will not occur with hadrons
containing $c$, and $\bar c$ (i.e.\ $D^\ast\bar D$, $D\bar D_s$ etc.) in
the FS\null. The latter has to be the case if rescattering phases were
all of short distance (SD) origin coming from the penguin
operators.\cite{bander,gerard2}

For FS such as $K\pi$, $\pi\pi$\dots wherein the two mesons
each have a non-vanishing isospin an interesting and important
possibility for CP violation arises due to the rescattering effects
caused by soft physics. For illustration, let us focus on $K\pi$. Then
one way that CPT maintenance can occur in the presence of nonvanishing PRA's in
the channel is for the PWD in the two possible modes of $B^-(B^0)$  to
cancel against each other so that\cite{atwood1}

\bea
\delta (K^-\pi^0) & = & -\delta (K^0\pi^-) \nonumber \\
& = & \delta(\bar K^0\pi^0) = -\delta(K^-\pi^+) \label{eq8}
\eea

\noindent where $\delta$'s denote the PWDs:

\be 
\delta(K^-\pi^0) \equiv \Gamma(B^-\to K^-\pi^0) - \Gamma(B^+\to
K^+\pi^0) \label{eq9}
\ee

\noindent etc. A simple calculation then gives

\be 
\delta(K^-\pi^0) = 2\sqrt{2} |V_u|\,|V_c|\,|A|\,|D| \sin\gamma
\sin\Phi \label{eq10}
\ee

\noindent where $V_u = V_{ub}V^\ast_{us}$, $V_c= V_{cb}V^\ast_{cs}$, $A$
is the $\Delta I=1/2$ penguin amplitude for $b\to s$, $D$ is the $\Delta
I=3/2$ tree amplitude, $\Phi$ is the strong rescattering phase,
$\Phi=\arg (DA^\ast)$ and $\gamma\approx \arg (-V^\ast_{ub})$ is the CP-odd
weak phase. The corresponding PRAs in these channels
become:\cite{atwood1}

\bea
{\rm PRA} (K^-\pi^0) & = {\rm PRA} (\bar K^0\pi^0) & = -2 {\rm PRA} 
(\bar K^0\pi^-) \nonumber \\
& & = -2{\rm PRA} (K^-\pi^+) \nonumber \\
&\;\;\; = \sqrt{2} r \sin\gamma \sin\Phi \label{eq11}
\eea

\noindent where $r= T_{K\pi}/P_{K\pi}$ for the tree versus penguin
amplitudes for $K\pi$. Setting $\sin\gamma=\sin\Phi =1$ for maximal
phases:

\bea
{\rm PRA}(K^-\pi^0)|_{\max} & = & {\rm PRA} (\bar K^0\pi^0) |_{\max} =
-2 {\rm PRA} (\bar K^0\pi^-) |_{\max} \nonumber \\
& = & -2 {\rm PRA} (K^-\pi^+) |_{\max}
= \sqrt{2} r \label{eq12} 
\eea

Assuming $r\sim.3$,
 the PRA in $K^-\pi^0$ and $\bar K^0\pi^0$ modes can be $\sim 42\%$
whereas that in $\bar K^0\pi^-$ and $K^-\pi^+$ is $\sim21\%$ (with the
opposite sign).\cite{atwood1}

We thus see that FS rescattering phases can cause appreciable direct CP
violating PRAs in exclusive modes such as $K\pi$. 

\subsection{Model Independent Test for EWP and/or New Physics in $K\pi$}

Many studies of EWP exist by now.\cite{fleischer} However, in their
numerical estimates these studies invoke many model dependent
assumptions and approximations. Model independent assessment of EWP
contributions are essential for possible application of these modes for
extraction of the unitarity angles and/or for the search for new
physics. 

Assuming isospin conservation and making no additional assumptions or
approximations allows us to arrive at a sum rule\cite{atwood1}

\be 
2|m_1|^2 - |m_2|^2 - |m_3|^2 + 2|m_4|^2 = 2|\bar m_1|^2 - \bar
m_2|^2 - |\bar m_3|^2 + 2|\bar m_4|^2 \label{eq13}
\ee

\noindent where $m_1$, $m_2$, $m_3$, $m_4$ are the amplitudes for $B$
decays to $K^+\pi^0$, $K^0\pi^+$, $K^+\pi^-$ and $K^0\pi^0$
respectively. 

A violation of this sum rule will be a model independent test for
EWP\null. If, on the other hand, such a model independent test shows
that EWP contribution is negligible, then the direct CP measurements in
the $K\pi$ modes would also become very useful for the extraction of the
angle $\alpha$ via a quadrangle construction\cite{nir} involving the
four modes of $(B^+,\bar B^0)$ and $(B^-,B^0)$ each. However, deducing
the weak phase $(\alpha)$ through this construction also requires, in
addition, time dependent oscillation studies of $(B^0,\bar B^0)$ to a
self-conjugate final state, i.e.\ $K^0_s\pi^0$.\cite{atwood1,nir}

\subsection{Decays to Two Vectors: Extraction of $\alpha$, $\gamma$ and
search for EWP.\protect\cite{atwood2}}

The decays $B$, $B_s\to 2$ vector final states can be especially
useful for determining the angles of the unitarity triangle, for model
independent tests of EWP and for the search for new physics. For our
method to work requires modes in which penguin and the tree are both
contributing. Specifically, the final states need satisfy the following
two conditions:

\begin{enumerate}

\item At least two decay amplitudes related by isospins must be involved.

\item The QCD penguin Hamiltonian must contributre only one isospin
amplitude to the FS. 

\end{enumerate}

Some illustrative examples are $B\to K^\ast \omega$, $K^\ast\rho$,
$\rho\omega$ and $\rho\phi$ and $B_s\to K^\ast\bar K^\ast$ and
$K^\ast\rho$. The method provides a model independent test (see below)
for EWP\null. Among the $B(B_s)$ modes mentioned above, those that are
least affected by EWP can be used for the extraction of $\alpha(\gamma)$

Once conditions (1) and (2) above are satisfied $B^-,\bar B^0$ decays to
a given FS can be used to write down a linear combination of the
amplitudes that contains only the phase of the tree. As in the previous
$K\pi$ case, to extract the weak phase $\alpha$ or $\gamma$, though,
time dependent oscillation studies of $B^0,\bar B^0$ to the
corresponding self conjugate FS are needed. 

Another important ingredient in this study is the correlations between
the decay distributions of the two vector mesons, or equivalently the
correlations of their polarizations.

Extraction of the information about the CKM phases proceeds along the
following key steps:

\begin{enumerate}

\item Following Gronau and London,\cite{gronau} isolate a linear
combination $(c_1u^h_1+c_2u^h_2)$ which contains only the weak phase of
the tree. Here $u^h_1$ and $u^h_2$ are the corresponding amplitudes for
$B^-$ and $\bar B^0$ decays; the superscript $h=1,0,{\rm+}1$ denotes the
helicity of the vector particles. This combination of the amplitude and
its CP conjugate will be related as follows:

\be 
(c_1u^h_1+c_2u^h_2) e^{-i\delta_T} = (c_1\bar u^h_1 + c_2\bar u^h_2)
e^{i\delta_T} \label{eq14}
\ee

where $\delta_T$ is the weak phase of the tree.

\item Angular distributions of the decay products of the two vector
particles is used to obtain the magnitudes of the helicity amplitudes
and the phases between the pairs of helicity amplitudes. 

%
%
%>>>>>>>>DA
%

\item Since the phase differences between different helicities for the
same FS can be obtained from the angular distributions, the set of three
equations eqn.~(\ref{eq14}) (one for each helicity) can now be regarded as
a set of three linear equations for the four remaining undetermined
phases, ie. one phase for each of the two modes and two conjugate modes. If
there is no EWP contamination, the equations will have a solution where
the phases are unimodular as expected.\cite{atwood2} On the other hand, if
such a solution cannot be found, it implies that eqn.~(\ref{eq14}) fails
so EWP contamination or a new physics contribution is present. This test
of EWP is completely model independent and is based only on isospin
conservation. 

\item Finally, as mentioned before, a time-dependent oscillation study
of $B^0$-$\bar B^0$ to the corresponding self-conjugate FS needs to be
done to fix the relative phase between the $B^0$ and the $\bar B^0$.
This enables the phases between all pairs of amplitudes to become known
so that the weak phase of the tree can thus be retrieved. 

\end{enumerate}

For $B\to K^\ast\omega$ the method should work especially well since the
color allowed EWP contribution, (resulting in $\gamma$, $Z\to\omega$),
carries the same weak phase as the strong penguin. Therefore, the
extraction of $\alpha$ is rather clean, suffering only from the
contamination of a color-suppressed EWP contribution which will most
likely be sub-dominant. In any case, the tests for EWP mentioned above
should be able to quantitatively monitor the extent of the
contamination.\cite{atwood2}

For $B\to K^\ast\rho$, since $\bar B^0\to\bar K^{\ast0}\rho^0$ is
susceptible to EWP contamination a slightly more involved strategy which
uses all four of the $K^\ast\rho$ modes but does allow a determination of
$\alpha$.\cite{atwood2}

Two other vector modes that may be useable for $\alpha$ extraction are
$B\to\rho\omega$ and $B\to\rho\phi$. For $B\to\rho\omega$ to work, it
will need to be demonstrated that EWP contamination is small as $B^0\to
\rho^0\omega$ receives color allowed EWP contribution from $\gamma$,
$Z\to\rho^0$.\cite{atwood2} For $B\to\rho\phi$, EWP contribution is not
expected to be a problem; however, its $Br$ is likely to be $0(10^{-6})$
whereas for $K^\ast\omega$, $K^\ast\rho$ and $\rho\omega$ the $Br$'s are
likely to be $0(10^{-5})$. 

The method can also be extended to determine $\gamma$ via $B_s\to
K^\ast\rho$ and $\bar K^\ast K^\ast$. 

\section{Chasing LUCY With Beauty: Search for a NSP Via $b\to sg^\ast$,
$B\to \eta^\prime X_s$ and Related Modes.}

There are good reasons to think that in addition to the CKM phase of the
SM, other CP violation phase(s) exist in nature. For one thing, it
appears difficult to account for baryon asymmetry in the universe 
with the CKM phase. Besides, extensions of the SM with new scalars,
fermions or gauge bosons almost invariably entail new phase(s). [LUCY is
a generic name for a non-standard phase (NSP)]. With the advent of
$B$-factories it is imperative to ask, how best can $B$-physics be used 
to search for a NSP\null. 

For a variety of reasons the penguin transition, $b\to sg^\ast$ (where
$g^\ast$ is an on or off-shell gluon) is a highly suitable probe of
LUCY\null. In inclusive processes the SM-CKM phase is likely to make
only a negligible contribution, so non-standard CP violating effects may
not get masked that easily. The fairly large $Br\sim 1$--2\% can be very
helpful. However a completely inclusive search for $b\to sg^\ast$ is
rather difficult. 

CLEO experimentalists have recently provided an ingenious clue for
harnessing an 
appreciable fraction of the $b\to sg^\ast$ signal. We are referring here
to an observation of an unexpectedly large inclusive signal of
$\eta^\prime$:\cite{browder}

\be 
Br(B\to \eta^\prime X_s) = [6.2\pm 1.6({\rm stat}) \pm 1.3 ({\rm
syst})^{+0.0}_{-1.5}({\rm bkg})] \times10^{-4} \label{eq15}
\ee

\noindent for $2.0<P\eta^\prime<2.7$ GeV\null. The magnitude of this
signal is about a factor of 5 bigger than estimates based on the SD
Hamiltonian. Of course, since the experimental discovery was first
reported a lot of theoretical effort has gone into understanding its
origin.\cite{atwood4,ali1} An interesting suggestion is that an appreciable
fraction of the $\eta^\prime$ signal originates from the coupling of the
$\eta^\prime$ to two gluons via the QCD anomaly\cite{atwood4} (see
Fig.~1). Recall that this anomalous coupling of the $\eta^\prime$
distinguishes the singlet $\eta^\prime$ from the flavored octet of
pseudo-goldstone bosons and renders it heavier than the other members of the
octet. The point is that the singlet axial current receives a
contribution from the gauge sector:

\be 
\partial_\mu j^\mu_5 = \frac{3\alpha_s}{4\pi} G_{\mu\nu} \tilde
G^{\mu\nu} + 2i\sum_{q=uds} m_q \bar q\gamma^5q \label{eq16}
\ee

Needless to say, quantitative estimates of this (or other) contributions
to the signal for $\eta^\prime X_s$ is very difficult. To begin with, the
mechanism was found to be roughly able to account for the observed
signal if one assumes that the $g^\ast-\eta^\prime-g$ form factor stays 
approximately constant with $q^2$ over the relevant range, $q^2$
being the 4-momentum of $g^\ast$. Since then 
further study by CLEO\cite{browder} of
the recoil mass distribution seems to give some additional support to
the anomaly idea.\cite{atwood4}

%
%
%
%
%
%+++++>>>>>>>>>>>>>>>>>>>>>>>Figure 1
%
%
%
%
%
\begin{figure}
\psfull
\begin{center}
\leavevmode
\epsfig{file=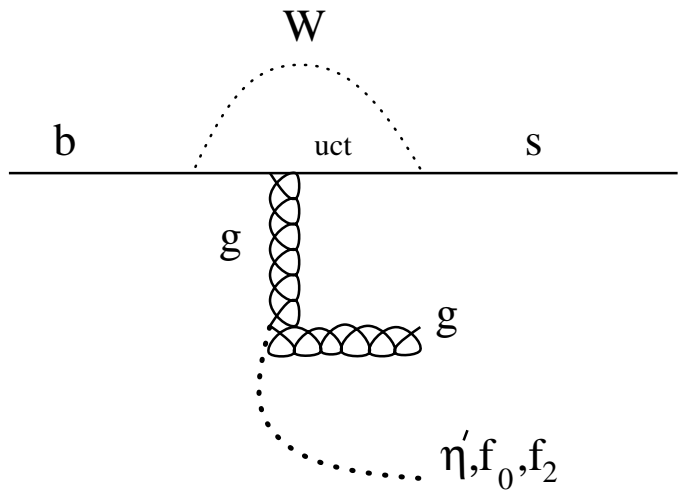,height=2.6 in}
\end{center}
\caption{\emph{
The penguin diagram giving rise to $b\to s g^*$ followed by
$g^*\to g +\eta^\prime$, $f_0$, $f_2$ or other such states.
}}
\label{brev_fig1}
\end{figure}

It should be emphasized that the $g^\ast-\eta^\prime-g$ form factor
introduces appreciable uncertainty into this calculation; the off-shell
gluon ($g^\ast$) carries a non-vanishing $q^2$, perhaps 0(a few
GeV$^2$). There is no good reason to think that this form factor would
be the same as the $\gamma^\ast-\pi^0-\gamma$ one,\cite{brodsky}
wherein it goes as $4\pi^2f^2_\pi/q^2$, with $f_\pi=130$ MeV\null. For
the gluonic-rich $\eta^\prime$, the analogous form factor may involve a
larger dimensionful parameter than $f_\pi$ 
e.g.\ constituent  quark mass and/or an
effective gluon or glueball mass.

This mechanism (see Fig.~1) is especially significant for searches for CP
violation\cite{atwood3,hou1} The key point is that the underlying
fragmentation process, $g^\ast\to g\eta^\prime$ renders the Feynman
amplitude for $b\to s\eta^\prime g$ for the inclusive $B\to \eta^\prime
X_s$ signal develop an imaginary part at the one loop. This (CP-even)
absorptive part is essential in driving observable CP violating
asymmetries. This is an interesting and perhaps an important difference
between $b\to \eta^\prime sg$ and the other large inclusive $b\to s$
process, namely $b\to s\gamma$ as, in perturbation theory, the latter
cannot develop a strong phase at one-loop. 

In the SM, though, as already mentioned the $b\to s\eta^\prime g$
amplitude receives only a negligible CP-odd phase. Therefore detection
of a largish CP asymmetry in this channel may be an indication for a
non-standard source of CP violation. 

For simplicity, we assume that the CKM phase in $b\to s$ penguin
amplitude is negligible. The $b\to sg^\ast$ effective vertex can be
parameterized as\cite{atwood3} (see. Fig.~2)

\bea
\Lambda^{bsg}_\mu & = & (v_t G_F/\sqrt{2}) \bar s_i T^a_{ij} [-iF(q^2) 
(q^2q_\mu-q_\mu\not q)L \nonumber \\
& & +(g_s/2\pi^2) m_b q_\mu \epsilon_\nu 
\sigma^{\mu\nu} G(q^2) R] b_j \label{eq17}
\eea

\noindent Here

\bea
F(q^2) & = & e^{i\delta_{st}} F_{SM} + e^{i\lambda_F} F_x \nonumber \\
G(q^2) & = & G_{SM} + e^{i\lambda_G}G_x \label{eq18}
\eea

\noindent where $\delta_{st}$ is the strong phase\cite{bander} generated
by the absorptive part resulting form the $c\bar c$ cut for
$q^2>4m^2_c$. To 
keep the discussion completely general we have introduced $\lambda_F$
and $\lambda_G$ as the CP-odd non-standard phases in the chromo-electric
and chromo-magnetic form factors.

%
%
%
%
%
%+++++>>>>>>>>>>>>>>>>>>>>>>>Figure 2
%
%
%
%
%
\begin{figure}
\psfull
\begin{center}
\leavevmode
\epsfig{file=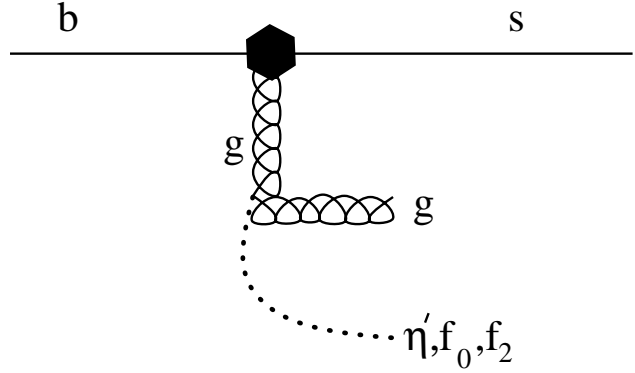,height=2.0 in}
\end{center}
\caption{\emph{
The contributions of non-standard physics to 
$b\to s g^*$ followed by
$g^*\to g +\eta^\prime$, $f_0$, $f_2$ or other such states.
The non-standard process is indicated by the hexagon.
}}
\label{brev_fig2}
\end{figure}

Although the inclusive $\eta^\prime$ signal is rather large, it is only
about 5\% of the total $b\to sg^\ast$. Besides, the $\eta^\prime$
detection efficiency is very low; at CLEO it is only about 5\% for the inclusive
signal.\cite{browder} It is, therefore, useful to ask, if the fragmentation reaction,
$g^\ast\to g\eta^\prime$ can lead to some other particle(s) in
addition to the $\eta^\prime$, an appreciable fraction of the time. 

Clearly the possible states that couple to two gluons have to have
$J^{PC}= 0^{-+}, 0^{++}$ and $2^{++}$. Some interesting examples
are\cite{atwood3}

\bea
0^{-+} & : & \eta(958), \eta^\prime(1440)\dots \nonumber \\
0^{++} & : & f_0(980), f_0(1370)\dots \nonumber \\
2^{++} & : & f_2(1270), f_2(1525)\dots \nonumber \\
{\rm Continuum} & : & \pi\pi, K\bar K, K\bar K\pi\dots \label{eq19}
\eea

\noindent One can also arrive at these candidate states by looking at
final states in radiative decays of the $\psi$.\cite{atwood3}

There are a couple of additional points worth keeping in mind:

\begin{enumerate}

\item At least in the case of $\psi$ decays\cite{atwood3} there appears
to be a close similarity between the states $(0^{-+})$ coupling to
$G\cdot \tilde G$ versus those $(0^{++})$ coupling to $G\cdot G$:

\bea
Br (\psi\to\phi\eta^\prime(958)) & = & (3.3\pm.4)\times10^{-4} \nonumber \\ 
Br(\psi\to\phi f_0(980))& = & (3.2\pm.9)\times 10^{-4} \nonumber \\
Br(\psi\to\omega\eta^\prime) & = & (1.67\pm.25) \times10^{-4}\nonumber \\ 
Br (\psi\to\omega f_0)& = & (1.4\pm.5) \times10^{-4} \label{eq20}
\eea

\item Inspection of $\psi$ decays\cite{barnett} also seems to show that
the signal for the $\eta(1440)$ may be comparable to that for the
$\eta^\prime$. 

\item A particularly interesting continuum of states is $K\bar K$ as it
leads to the overall signal of $B\to K\bar K+X_s$ (or $B\to K\bar K+X_d$)
which may be quite distinctive. Once again, inspection of $\psi$ decays
suggests that the ratio of $Br$'s: $(g^\ast\to K\bar K+g)/(g^\ast \to
\eta^\prime+g)$ could be around one or even somewhat bigger. 

\end{enumerate}

Calculation of the PRA proceeds along well known lines.
Fig.~3(a) and 3(b) show the PRA as a function of the recoil mass ($\mrec$) for
states of different $J^{PC}$. Fig.~3(a) is the case when beyond the SM
(BSM) physics 
contributes only 10\% to the production rate and Fig.~3(b) is assuming that
50\% of the rate is due to new physics. In both figures we have set the
CP-odd NSP to be maximal (i.e.\ $\lambda_F$, $\lambda_G = \pi/2$). 
These figures also
show separately the cases when the chromoelectric form factor dominates
(and the chromomagnetic is zero) and vice-versa. We see that PRA tend to
be bigger (by 50--100\%) for the case when chromoelectric dominates.
Notice that the asymmetries are $\sim 12$--34\% in Fig.~3(b) and in
Fig.~3(a), which is for the case when NSP is contributing only 10\% to the
rate, the asymmetries are still appreciable $\sim 8$--17\%. This is
quite remarkable as it would be virtually impossible to detect the
presence of the new physics, if BSM physics contributes only 10\% to
the rate, by comparison of the measured rate with
theoretical expectation as the rate calculations are extremely
unreliable. Search for CP violation can clearly be extremely worthwhile
and decisive.

%
%
%
%
%
%+++++>>>>>>>>>>>>>>>>>>>>>>>Figure 3+4 together
%
%
%
%
%
\begin{figure}
\psfull
\begin{center}
\leavevmode
\epsfig{file=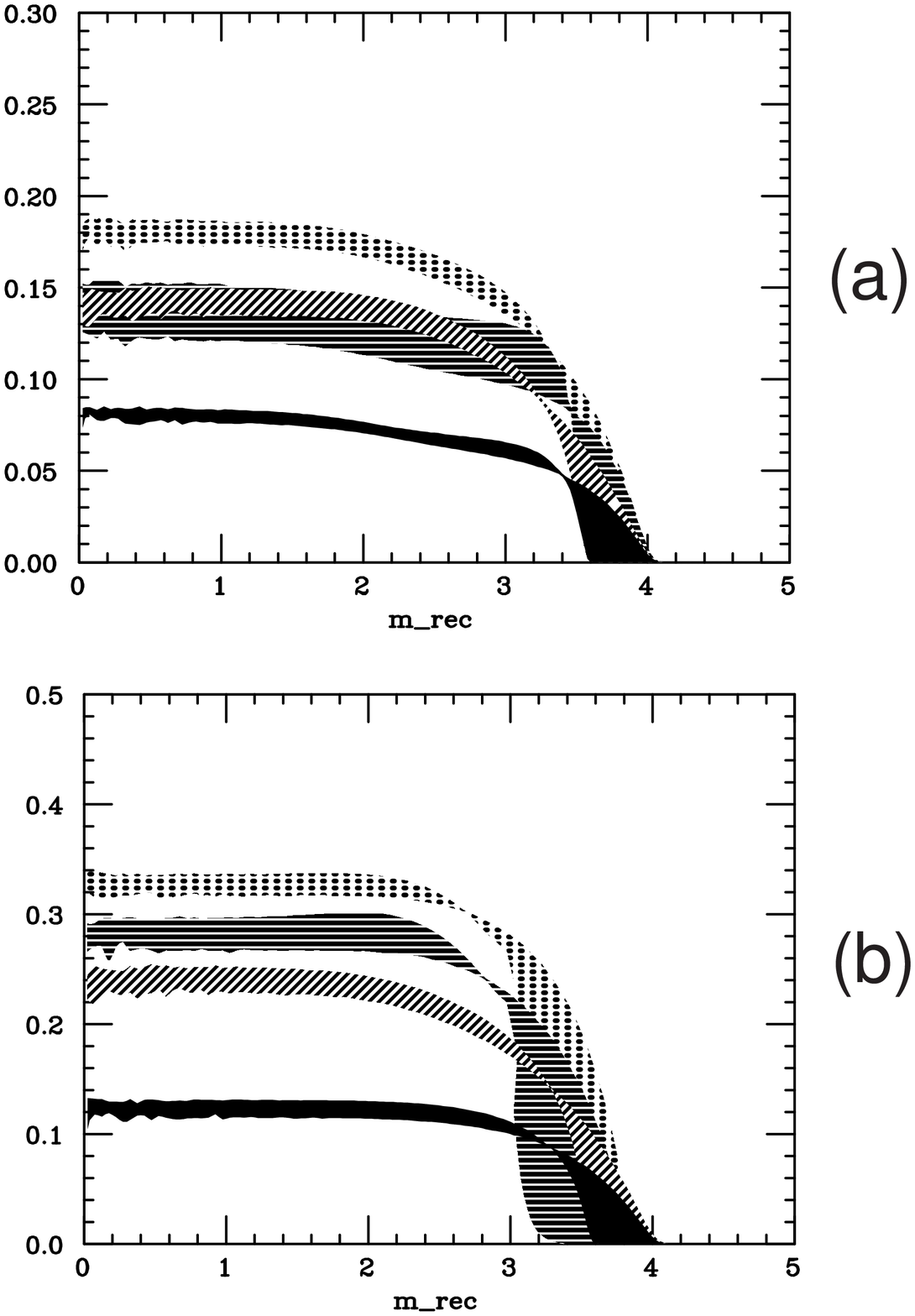,height=5 in}
\end{center}
\caption{\emph{
(a)~$\pra$
versus $\mrec$ assuming non-standard physics (NSP) contributes
$\sim 10\%$ to the rate for each state.  Also, $\sin\lamF=\sin\lamG=1$ is
used. The black shading shows the $\pra$ for $b\to s+g+ 0^{-+}$ assuming that
$\fx =0$ and taking $m_{0^{-+}}$ to vary from $958 MeV$ to  $1725 MeV$. The
horizontal striped region is the $\pra$ for the same $0^{-+}$ states, now
assuming that $\gx =0$.  Note that
the region indicated by the diagonal stripes shows the $\pra$ for
$b\to s+g+ 0^{++}$ assuming that $\fx =0$ with $m_{0^{++}}$
ranging from $980 MeV$
to  $1710 MeV$; the dotted region is the $\pra$ for the $0^{++}$ assuming 
that
$\gx =0$.
(b)~$\pra$
versus $\mrec$ assuming NSP contributes $\sim
50\%$ to the rate for each state. The black shading ($\fx =0$) and the
horizontal striped ($\gx =0$) are for $0^{-+}$ states as in Fig.~3(a). Diagonal
striped ($\fx =0$) and the dotted ($\gx =0$) regions are now for
$b\to s+g+2^{++}$
with $m_{2^{++}}$ from $1270$ to $2300MeV$. See also caption to Fig.~3(a).
}}
\label{brev_fig3}
\end{figure}

\newpage

\section{$b\to dg^\ast$, $B\to \eta^\prime X_d$ and Related Modes for
Direct CP violation due SM.\protect\cite{atwood6}}

The penguin $b\to d$ transition interfering with the tree $b\to u\bar u
d$ can be a rich source of direct CP violation due to the SM-CKM phase.
Specifically it is important to try to use $b\to dg^\ast$. Once again
$g^\ast\to g\eta^\prime$ and other fragments discussed in the preceding
section can be used. While the $Br$ for $B\to \eta^\prime X_d$ is
expected to be somewhat smaller ($\sim7\times10^{-5}$) than $B\to
\eta^\prime X_s$, the PRA driven by the CKM phase are expected to be
appreciably bigger for the $B\to \eta^\prime X_d$ case compared to that
for $B\to \eta^\prime X_s$. 
%
%
% >>>>>>>>>>>DA
%
%
For values of the CKM matrix 
which match current constraints,
PRA $\sim O(12\%)$ for $B\to
\eta^\prime X_d$. The asymmetries for $B\to \eta(1440)+X_d$,
$f_0(980)+X_d$, and $f_2(1270)+X_d$ are quite similar but somewhat 
bigger.\cite{atwood6}

%
%
%
%
%
%+++++>>>>>>>>>>>>>>>>>>>>>>>Figure 5
%
%
%
%
%
%\begin{figure}
%\psfull
%\begin{center}
%\leavevmode
%\epsfig{file=brev_fig4.eps,height=2.6 in}
%\end{center}
% 
%\label{brev_fig5}
%\end{figure}
%

\section{Two Body Decays: $b\to qM$ for Direct CP Studies and for
Looking for EWP.\protect\cite{atwood5}}

The importance of the two body decays of the $b$-quark is due to the
fact that their theory is relatively simpler and many of them have
relatively clean experimental signatures. To the extent that the
spectator approximation works, these 2-body decays at the quark level
end up materializing in quasi-two-body (QS2B) decays of the $B$-meson. Indeed,
a famous recent example is $B\to \eta^\prime X_s$ although from the
point of view of the theory, the $\eta^\prime$ case is somewhat of a
special one due to the unique relation of the $\eta^\prime$ to the QCD
anomaly. As far as experimental detection goes, the inclusive
$\eta\prime$ does serve an excellent example for the semi-inclusive
decays $B\to MX$ that are under discussion in this section. As in the
case for the $\eta^\prime$, the meson $M$ originating from the 2-body
decay $b\to Mq_f$ tends to obey two body kinematics and therefore has
$E_M\sim 2.5\pm.3$ GeV\null. The quark $q_f$ leads to the debris of
mesons with (1) relatively low average multiplicity $\sim3$, (2) total
energy $\sim 2\pm.3$ GeV and (3) with negligible total transverse
momentum (to $\vec p_M$ in the rest frame of $B$).\cite{atwood5}

Theoretical calculation for the QS2B decays are simpler than the
traditional exclusive mode such as $B\to K\pi$, $\rho\pi\dots$ A key
difference is that the latter reactions require knowledge of exclusive
form factors, e.g.\ $\langle B|J^h_1|\rho\rangle$ or $\langle
B|J^h_2|K\rangle$\dots where $J$'s are the appropriate quark level
currents. The QS2B reactions bypass the need for these form factors as
they represent 
a sum over the exclusive channels. Indeed, as a rule, the
$Br$ for the QS2B tends to be bigger compared to the exclusive two body
modes, in some cases by factors of $\sim 5$--10. In instances where the
formation of $M$ via the QS2B reactions is color suppressed whereas $M$
can be made in a color allowed manner with the participation of the
spectator for the corresponding exclusive two body mesons reactions,
then the latter can have a bigger $Br$ compared to the QS2B\null. This
may well, for example, be the case for $B^\pm\to \omega\rho^\pm$ versus
$b\to \omega d$. 

For these QS2B decays the CP-even FS interaction phase needed for
generating the CP violating PRAs arises from the penguin Hamiltonian.
Indeed since these are semi-inclusive reactions the calculations of the
FS phase originating from the absorptive part of the penguin graph is
close in spirit to the original quark level calculation.\cite{bander} It
is quite likely that the FS rescattering effects due to LD or soft
(non-perturbative) physics are much smaller in these inclusive processes
than in the corresponding exclusive channels.\cite{atwood1}

A quick look at the Table shows that the PRA may be quite large in
several channels,\cite{atwood5,datta} e.g.\ $M=K^\ast, K, \rho,\pi\dots$
Also given in the Table is $N^{3\sigma}_B$, the number of $B$ mesons needed to 
obtain a $3-\sigma$ signal for the PRA, as well as an estimate of the EWP 
contribution. 
While it would be interesting and important to search for these PRA it
would be very worthwhile even if the $Br$'s of some of these channels
are measured. The resulting input from experiment could be valuable in
fine tuning the calculational parameters of this whole class of
reactions. In this context, it is also worth noticing that the recoil
spectrum (see Fig.~4) for these reactions can be calculated in
much the same way as for the $B\to\gamma X_s$\cite{ali2} (or $B\to\eta
X_s$\cite{atwood4}). This 
means that the input parameters for the recoil spectrum can be refined by
using data from all of these reactions.

%%%%%%%%%>>>>>>>>>>>>>>DA
%
%
%    revised table:::::::::::::>>>>>>>>>>>>>>>>

\begin{table}[ht]
\caption{Some modes of interest; $Br$, PRA and $N^{3\sigma}_B$ along with
EWP contributions (to color allowed channels only) are shown. 
Note $\gamma\equiv\arg (-V^\ast_{ub}V_{ud}/V^\ast_{cb}
V_{cd})$. Note that the $Br$ column does not include contributions
from EWP.\label{tabone}} 
\begin{center}
\begin{tabular}{l|c|c|c|c}
\hline
Mode & $Br$ & $|{\rm PRA}|/\sin\gamma$(\%) 
& $N^{3\sigma}_B\sin^2\gamma \epsilon_{eff}/10^6$ & $Br$ due to EWP \\ 
\hline
$\pi^-u$           & $1.3\times10^{-4}$   & 8   & 12 &    \\
$\rho^-u$          & $3.5\times10^{-4}$   & 8   &  4 &    \\
$\pi^0d$           & $2.4\times10^{-6}$   & 36  & 28 &    
$4.7\times10^{-8}$ \\
$\rho^0d$          & $5.9\times10^{-6}$   & 38  & 10 &    
$1.3\times10^{-7}$ \\
$\omega d$        & $5.8\times10^{-6}$   & 39  & 10 &     
$7.0\times10^{-9}$ \\
$\phi d$           & $2.3\times10^{-7}$   & 0   &\  &     
$7.0\times10^{-9}$ \\
$K^0s$             & $2.5\times10^{-6}$   & 5   & 1200 &    \\
$K^{0^\ast}s$      & $2.9\times10^{-6}$   & 16  & 120 &    \\
$D^-c$             & $1.7\times10^{-3}$   & 2   & 17 &    \\
$D^{\ast-}c$       & $2.2\times10^{-3}$   & 2   & 13 &    \\
& & & \\
$K^-u$             & $2.9\times10^{-5}$   & 33  & 3 &    \\
$K^{\ast-}u$       & $5.1\times10^{-5}$   & 51  & 1 &    \\
$\bar K^0d$        & $2.0\times10^{-5}$   & 1   & 3000&    \\
$\bar K^{0^\ast}d$ & $2.6\times10^{-5}$   & 3   & 540 &    \\
$\pi^0s$           & $9.8\times10^{-8}$   & 0   &\  &$1.6\times10^{-6}$ \\
$\rho^0s$          & $2.5\times10^{-7}$   & 0   &\  &$4.3\times10^{-6}$ \\
$\omega s$       & $1.3\times10^{-6}$   & 0   &\  &$4.7\times10^{-7}$ \\
$\phi s$           & $6.3\times10^{-5}$   & 0   &\  &$4.7\times10^{-7}$ \\
$D^-_sc$           & $4.2\times10^{-2}$   & 0.1   & 300  &    \\
$D^{\ast-}_sc$     & $5.3\times10^{-2}$   & 0   & 300  &    \\ 
\hline
\end{tabular}
\end{center}
\end{table}
%

%
%
%
%
%
%+++++>>>>>>>>>>>>>>>>>>>>>>>Figure 6
%
%
%
%
%
\begin{figure}
\psfull
\begin{center}
\leavevmode
\epsfig{file=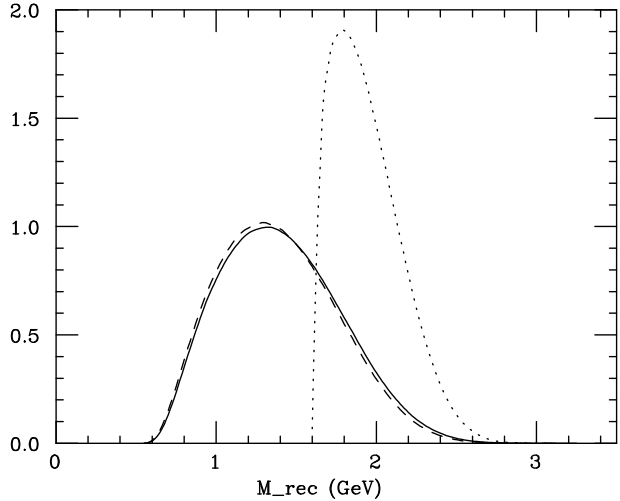,height=2.6 in}
\end{center}
\caption{\emph{
The
normalized
recoil spectra for the
quasi-twobody decays, $b\to\pi^-u$ (solid), $K^{\ast-}u$ (dashed) and
$D^-c$ (dotted) are shown.
}}
\label{brev_fig6}
\end{figure}

Finally, we draw attention to the fact that $B^0\to \rho^0X_s$ and
$B^0\to\pi^0 X_s$ appear highly suited for searching for
EWP.\cite{atwood5} This can be qualitatively understood as the EWP
contributions are color allowed whereas the others are color suppressed.
The contribution from the tree graph is also CKM suppressed. The EWP
appears to be the dominant contributor\cite{atwood5} with 
$Br(B^0\to\rho^0X_s)\sim4\times 10^{-6}$ and $Br(B^0\to\pi^0 X_s)\sim2
\times 10^{-6}$. 
%
%
%>>>>>>>>>>>>>>>>>>>>DA
%
The same comments may also be true for the related reactions,  
$B^+\to \rho^0 X_s$,
$B^+\to \pi^0 X_s$, except that in this case there is a possible background 
from the tree process $\bar b\to u\bar u \bar s$ wherein the 
$\bar u$ can combine with the spectator to form 
the $\rho^0$ or $\pi^0$ in a color allowed manner. 
Although most of the 
background from the tree process would be expected to give rise to lower 
momentum $\rho^0$ or $\pi^0$, one can estimate that the 
high momentum component of the $\rho^0$ or $\pi^0$ energy spectrum 
can produce a signal
comparable to 
the EWP contribution.

\section*{Acknowledgements}

We are grateful to the organizers of the International Workshop, and
especially to George Hou, for the very warm 
and gracious hospitality. This research was
supported in part by US DOE Contract Nos.\ DE-FG01-94ER40817 (ISU) and
DE-AC02-98CH10886 (BNL).

\end{document}